\documentclass[conference]{IEEEtran}
\IEEEoverridecommandlockouts
\usepackage{cite}
\usepackage{amsmath,amssymb,amsfonts}
\usepackage{algorithmic}
\usepackage{graphicx}
\usepackage{textcomp}
\usepackage{xcolor}
\usepackage{authblk}

\def\BibTeX{{\rm B\kern-.05em{\sc i\kern-.025em b}\kern-.08em
    T\kern-.1667em\lower.7ex\hbox{E}\kern-.125emX}}
\begin{document}


\title{Towards Learning How to Properly Play UNO with the iCub Robot\\
}

\author[1]{Pablo Barros}
\author[1]{Stefan Wermter}
\author[2]{Alessandra Sciutti}
\affil[1]{Knowledge Technology, University of Hamburg, Hamburg, Germany \authorcr Email: {\tt \{barros, wermter\}@informatik.uni-hamburg.de}\vspace{1.0ex}}
\affil[2]{Cognitive Architecture of Collaborative Technologies Unit, Istituto Italiano di Tecnologia, 
Genoa, Italy \authorcr Email: {\tt alessandra.sciutti@iit.it} \vspace{-1.0ex}}




\maketitle

\begin{abstract}
While interacting with another person, our reactions and behavior are much affected by the emotional changes within the temporal context of the interaction. Our intrinsic affective appraisal comprising perception, self-assessment, and the affective memories with similar social experiences will drive specific, and in most cases addressed as proper, reactions within the interaction. This paper proposes the roadmap for the development of multimodal research which aims to empower a robot with the capability to provide proper social responses in a Human-Robot Interaction (HRI) scenario.
\end{abstract}


\section{Introduction}

Our capabilities of both perceiving and reacting to the affective behavior of other persons are fine-tuned based on the observed social response of our interaction peers. We usually perceive how others are behaving towards us by reading their affective behavior through the processing of audio/visual cues \cite{lang1997motivated}. We are then able to provide an affective response, either by modulating our voice, using facial expressions, producing body movements, or realizing a specific action within the interaction context. The response is usually modulated by a contagion effect \cite{hatfield1993emotional}, where we are impacted by and replicate the perceived affective state of others, or by reacting with a proper social response \cite{hao2017enhancing}, which is accepted and expected by the interaction peers. Deciding the proper response is a learning process modulated by the interpretation of the impact that our behavior causes within the interaction context \cite{ochsner2008cognitive}. The learning association between perception and action configures a long-term adaptation mechanism which updates how we perceive and respond to others over a lifelong span. Adapting such a mechanism into a robotic platform would allow it to provide socially acceptable responses when in interactive social scenarios, and would benefit greatly the acceptability and human-likeliness of companion and assistive robots \cite{crossman2018influence}.


Adapting this perception-action mechanism into a social robot demands the development and integration of two mechanisms: an adaptive emotion appraisal, which is updated over time towards novel persons, and a behavior generation which learns based on the perception of the social impact that a certain action caused within the interaction.

The state-of-the-art in emotion recognition for social robots was much benefited with the recent advents of deep neural networks \cite{chen2018softmax}. Such models can learn, implicitly, how to represent high-level stimuli, such as facial expressions and prosody in speech. Such capability allows these models to represent multimodal emotion expressions efficiently and robustly, and they currently post as the state-of-the-art solutions in different emotion recognition tasks. When deployed to real-world scenarios, in particular, involving social robots, however, these solutions tend to fail, mostly due to their inability to adapt quickly to how specific persons express emotions \cite{sariyanidi2014automatic}. Deep learning models have millions of different parameters which need to be updated to depict different affective characteristics. Given the subjectivity on perceiving and labeling emotion expressions, to update these models to learn the specific characteristics of how a person expresses emotions without a large number of training samples of that specific person posts as a machine learning challenge.

Generating behavior based on affective perception has been deeply studied by the social robots area \cite{hirokawa2018adaptive}. Most of the proposed models, however, employ an emotion contagion strategy which usually reproduces the perceived facial expressions or body movement \cite{van2018generic}, or by using a simple decision tree on how to behave when a certain emotion was perceived \cite{tuyen2018emotional}. Such solutions are ineffective when dealing with complex interaction scenarios, such as interactions which are based on empathetic responses, very common to conversational robots and robots are driven by social responses \cite{leite2013social}. As soon as the robot has to provide a response which was not scripted by pre-defined decision mechanisms, it will usually fail to maintain an human-level of engagement with the users \cite{sidner2004look}, which was observed to reduce its applicability when interacting with children or elderly citizens in therapy or health-care scenarios \cite{chandrasekaran2015human}. One of the most common machine learning solutions to generate robot behavior is to provide a learning mechanism based on reinforcement learning. Usually, in such learning scheme, a robot reads the feedback from the environment and use it in a reward function to modulate its learning. The advantage of using reinforcement learning is that the robot can learn by itself, with little or no human supervision, by measuring the effect that its actions caused in the environment. When deployed in Human-Robot Interaction (HRI) scenarios, however, reinforcement learning strategies are usually not effective as they demand several thousands of examples to learn associations between perception and actions. As in most social interaction scenarios the robot interacts with a human, thousands of interactions would be necessary before the robot learns meaningful responses \cite{churamani2018learning}, reducing its applicability in real-world scenarios.

This position paper focuses on a multi-subject interaction scenario where different humans play a competitive card game, known as UNO, with a robot. The robot will be an active part of the interaction and will play against humans. As such, the robot will perceive the individual and group affective behavior and the game status and derive an action which is composed of behavioral response (facial expression and body movement), and a game-related decision (e.g. play a card or collect a card).


\begin{figure}[t]
\begin{center}
   \includegraphics[width=0.55\linewidth]{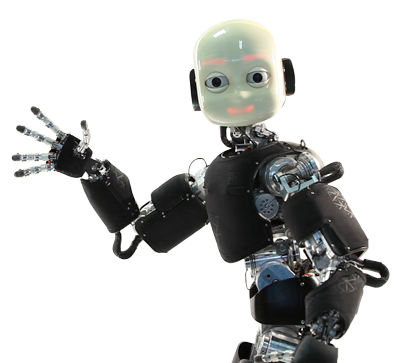}
\end{center}
   \caption{The iCub humanoid robot. We will use it in our proposed scenario due to its social capabilities, both for the perception of audio/visual cues and the generation of facial expressions and body movement.}
\label{fig:icubRobot}
\end{figure}

We also describe in this paper the first steps towards a hybrid neural architecture to be embedded into an iCub robot\cite{metta2008icub}, illustrated in Figure \ref{fig:icubRobot}. We drive the development of the proposed framework based on two directions: an adaptive perception neural network which uses an unsupervised affective memory mechanism to overcome the problems of online learning in deep neural networks to adapt towards how specific persons express emotions, and to model an intrinsic affective state; and an interactive reinforcement learning strategy based on learning the social and contextual impact of a certain action by measuring the affective responses of the players, avoiding the necessity of active human interaction in the learning loop.

This paper details the road-map on how the individual models that compose the framework will be developed and integrated. We also describe, and justify, in details the evaluation scenario and protocol to obtain a set of objective evaluation measures to evaluate the impact of our proposed framework.

\section{Playing UNO with a Robot}

The iCub robot is one of the most advanced platforms for Human-Robot Interaction (HRI). It was evaluated in different interaction scenarios and it provides a high-level of engagement, acceptability, and reliability when interacting with different persons. These are important characteristics to improve the perception of the robot as an active social member in our scenario, which we assume it will be an important characteristic to elicit natural affective behavior. 

In our proposed interaction scenario, the iCub is playing the card game named UNO with three other persons. In this game, each player receives a set of cards, with different colors, numbers, and possible actions. Starting from a pre-selected card, the players have to discard a card which matches the color or number of the pre-selected card. The game happens in turns. The first player which discard all the cards at hand wins the game. The action cards help to improve the competitiveness of the game by providing specific game-change events, such as ``make the next player collect 4 cards", or `` invert the playing order". In our scenario, the iCub knows the rules of the game and can do a set of pre-defined actions such as check its cards, discarding and collecting cards, calling ``UNO", and passing its turn. The robot can also display certain affective responses, using different body movements and facial expressions.

\begin{figure}[t]
\begin{center}
   \includegraphics[width=0.80\linewidth]{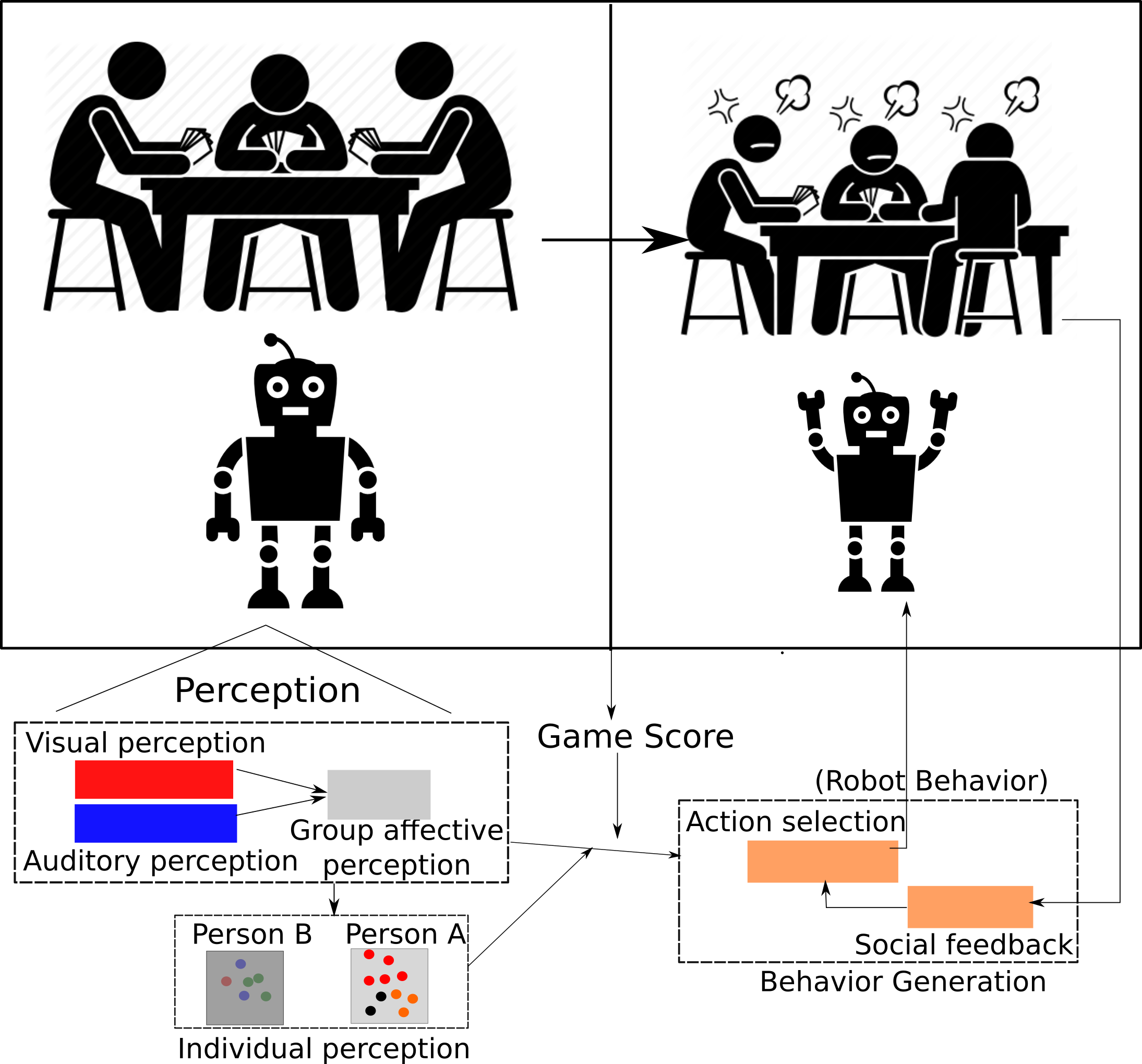}
\end{center}
   \caption{Our competitive game scenario involving three persons and the iCub robot. The robot will read audio/visual cues of the persons and the game context. The iCub will produce an action composed of a game-related action, and affective behavior. The impact of the produced action on the game and the persons is read and used to identify if the action was proper, given the situation, and to modulate the learning of novel actions.}
\label{fig:scenario}
\end{figure}

While playing, the robot will observe the individual affective behavior of the other humans and will decide which game-related action to take and which affective expression it will display. The UNO card game was chosen as it is a relatively controllable scenario, where each player has its turn to make an action, a highly competitive game in which trying to identify, or even manipulate, the other players' affective behavior is an important game strategy, and, most importantly, the game provides several turns of context-dependent interaction as illustrated in Figure \ref{fig:scenario}.


The iCub must take into consideration how the three persons are behaving while playing to be able to react specifically and in an appropriate manner to each of them. The appropriate manner will be selected based on different desired targets: the iCub must win the game (maximizing the game strategy), it must improve the engagement of the other players (maximizing the generation of affective behavior), or it must provide a playing behavior which is accepted by the subjects as human-like (maximizing both affective behavior and game strategies).

As each person expresses emotions in a highly individual manner, the iCub must adapt its perception to each of them. But, as they pose as a group, it also must take into consideration how the affective behavior of each person impacts the reading of the emotional state of the entire group. The iCub must take into consideration the gradual evolution of the affective states of each player within the game, and associate it with the temporal change within the game context. 


We derived this scenario from common real-world applications for companion and tutoring robots used in elderly care centers and interaction with children. Our scenario, however, increases the complexity of the interaction by providing an unrestricted turn-taking initiative, where the persons will be instructed to act as natural as possible.  To provide a real-time response with the iCub is of much importance, as it must be perceived as an active and autonomous player. Also, the interplay between the individual perception and the group affective behavior is very important for grounding the produced actions and for the developing of different game-related strategies.

\section{Towards a Cognitive Framework}

\subsection{Affective Modeling}

Recently, we proposed a novel hybrid neural framework to deal with the problem of adapting deep neural models towards specific characteristics of perceiving emotion expressions in an online manner \cite{barros2016developing}. A series of pre-trained convolution channels were used to represent facial expressions, body movement, and prosody on the speech from one person. The GWR network received the multimodal representation as input, and learning, online, how to cluster the sensory stimuli into different affective concepts. The GWR adapts to the perceived stimuli by creating or removing neurons and it can learn, quickly, how to model novel information. Each neuron represents a prototype emotional concept, and are used to represent affective expressions. We explored the use of individual GWRs to learn prototype emotional concepts from individual persons and used them as specific affective memories which represent how that specific person expresses emotions \cite{barros2017self}. The affective memories were evaluated on an emotion recognition task and were shown to be a state-of-the-art model on recognizing spontaneous emotion expressions.   

To deal with the proposed scenario, we will take inspirations on the affective memory networks. We must devise a novel affective memory model which can deal with the temporal changes within the interaction, in particular, able to integrate the perceived stimuli asynchronously. The model will process facial expressions, body movement, and speech signals from all persons in the scenario,  which are all expressed within different contextual times. To deal with this difference we will investigate the use of gated recurrent units (GRUs) \cite{cho2014learning} which will contribute to a time-compressed representation of an expression. A general attention mechanism will also important to avoid that intense emotional changes are neutralized by long periods of displaying the same affective behavior. To address this problem we will integrate the concepts of local and global attention to the convolution channels of the model. Also, we will investigate the use of recurrent pooling layers will be used to highlight specific periods of the input stimuli which contribute more to emotion recognition. 

To provide a general affective analysis of the situation, we will integrate the individual affective memories into one emotional read. We will investigate the use of recurrent gamma connections \cite{Parisi2018} within a GWR to model the impact that each person will have in the general emotional representation. The GWR, once again, will provide the model with an online association learning mechanism, which will be able to depict, in an online manner, how both persons are behaving based on their affective information. 

 It is important to ground the proposed model on existing literature on emotion processing. This allows a direct comparison with humans not only on the performance level but also on behavioral aspects. Thus, the design of the novel learning rules, the topological and conceptual aspects of the model will be inspired by behavioral knowledge on how humans perceived emotion expressions.

\subsection{Modulating Game Strategies with Social Reactions}

To learn how to provide proper behavior, we will develop and evaluate a reinforcement learning model based on actor-critic neural networks. The actor-network will learn how to associate the current group perception state of the robot and the affective memory related to the person that the robot is currently aiming to play with, with a set of game and affective behavior actions. 

By perceiving the reaction of the players to the selected action, and the change in the game context, the iCub will maximize the production of a proper reaction. The critic network will learn how to maximize the actor-network output based on the desired target: a reward function based only on maximizing the game score of the iCub will produce a set of behaviors which will drive the iCub to win the game; by providing a set of reaction that maintain the group affective state positive, will direct the iCub to produce behaviors which seem as enjoyable by the other players; and creating a reward function based on both improving the game score of the iCub and maintaining a positive affective behavior on the group, will develop a certain game strategy that we assume it will be perceived as more human-like.




To maximize the learning strategy based on the few data points collected in each interaction, we will investigate the use of predictive stimulation \cite{nagai2019predictive} to provide a continual learning mechanism for the actor-critic network. We will collect videos of persons playing the game, and use it to create a replay memory pool, which integrates perception-action-reaction associations.  Different modulation techniques, in particular, based on the interplay between exploration and exploitation will be devised to avoid that the model overfits towards the memory replay.

\section{Evaluation protocol}

We will evaluate the proposed model using two strategies: one offline, which will support the topological development of each model, and one with real persons, which will enforce the real-time processing properties demanded by our scenario.

The offline training will be important for the development of novel learning rules and hyperparameter optimization of the model. We will simulate the game using the strategies collected from the recorded data, and the iCub simulator with a set of pre-defined game strategies. We will evaluate the individual modules of the proposed model, as well as different integration strategies. For this evaluation, objective measures such as the accuracy on recognizing emotion expressions, the concordance between the model and external human observers and the processing time will be maximized.

The real-world evaluation will involve the described scenario. Ideally, we will have multiple sessions with the same participants happening over a month to capture the important characteristics of the affective behavior of each participant. We will shuffle each group of participants during the sessions to avoid perception bias associated always with the same group of participants. The expected behavior is that the robot learns how to play the game based on the three different goals, and devise specific playing characteristics for each person. After each interaction, the participants will fill out different HRI-related questionnaires. We are in particular interested in the Asch`s personality impression questionnaire \cite{asch1946forming}, which will give us the persons´ assessment of how the robot is behaving, and the Godspeed questionnaire \cite{bartneck2009measurement}, which will help us to understand the persons’ general opinion about the impact of the robot in the interaction. Combining these results with the objective measures of the offline evaluation will give us a general perspective on how the proposed model impacts the general acceptance of the robot in a competitive scenario.

\section{Conclusion}

The action-perception cycle of perceiving emotion expressions, and responding with socially accepted actions is one of the most important aspects of human-human natural communication. In this paper, we describe our intentions to develop a hybrid neural framework for learning proper affective responses with an iCub robot in a competitive game scenario. The framework will comprise of two modules, the first one implementing deep and unsupervised neural networks for learning how to represent the group affective behavior from the specific characteristics of individual persons. The second module uses an actor-critic reinforcement learning neural network to provide actions, based on the perceived affective behavior. The impact of a chosen actions, based on three different desired targets, will be used as learning modulation for the model. We also present a series of evaluation strategies on how to provide a balance between objective and subjective analysis to improve the robustness of the model and to assess its impact on our HRI scenario.

\section*{Acknowledgment}
The authors gratefully acknowledge partial support from the German Research Foundation DFG under project CML (TRR 169). Alessanda Sciutti was supported by a Starting Grant from the European Research Council (ERC) under the European Union’s Horizon 2020 research and innovation
programme, G.A. No 804388, wHiSPER.

\bibliographystyle{plain}
\bibliography{bib}

\end{document}